\documentclass[aps,prb,twocolumn,floatfix,showpacs,citeautoscript,longbibliography,hyperlinks]{revtex4-1}

\usepackage{graphicx}
\usepackage{dcolumn}
\usepackage{bm}
\usepackage{bbold}
\usepackage{mathtools}
\usepackage{color}
\usepackage{transparent}
\usepackage[colorlinks=true,citecolor=blue]{hyperref}

\DeclarePairedDelimiter{\floor}{\lfloor}{\rfloor}

\begin{document}

\title{Electron waiting times of a periodically driven single-electron turnstile}
\author{Elina Potanina}
\author{Christian Flindt}
\affiliation{
Department of Applied Physics, Aalto University, 00076 Aalto, Finland}
\date{\today}

\begin{abstract}
We investigate the distribution of waiting times between electrons emitted from a periodically driven single-electron turnstile. To this end, we develop a scheme for analytic calculations of the waiting time distributions for arbitrary periodic driving protocols. We illustrate the general framework by considering a driven tunnel junction before moving on to the more involved single-electron turnstile. The waiting time distributions are evaluated at low temperatures for square-wave and harmonic driving protocols. In the adiabatic regime, the dynamics of the turnstile is synchronized with the external drive. As the non-adiabatic regime is approached, the waiting time distribution becomes dominated by cycle-missing events in which the turnstile fails to emit within one or several periods. We also discuss the influence of finite electronic temperatures. The  waiting time distributions provide a useful characterization of the driven single-electron turnstile with complementary information compared to what can be learned from conventional current measurements.
\end{abstract}

\maketitle

\section{\label{sec:level1}Introduction}

Dynamic single-electron sources are expected to play a central role in future quantum technologies based on the accurate emission of single electrons into quantum electronic circuits.\cite{[{See the special issue edited by }] Splettstoesser2017} For example, in a quantum information processor working with a fixed clock cycle, the periodic emission of single electrons will be important for synchronized many-particle operations.\cite{Bocquillon2014} Moreover, dynamic single-electron emitters may generate quantized electrical currents that are given exactly by the driving frequency times the electronic charge.\cite{Giblin2012,Pekola2013} Dynamic single-electron emitters have been realized in several experiments based on charge pumps \cite{Pothier1992,Ono2003,Blumenthal2007,Fujiwara2008,Kaestner2008,Jehl2013,Rossi2014,Ubbelohde2014,Fricke2014}, turnstiles \cite{Geerligs1990,Kouwenhoven1991,vanZanten2016} and mesoscopic capacitors,\cite{Feve2007,Bocquillon2013} or by applying lorentzian-shape voltage pulses to a contact.\cite{Dubois2013,Jullien2014}

The accuracy of the emitters can be characterized by measuring the low-frequency fluctuations of the electrical current.\cite{Blanter2000,Kaestner2015,Maire2008} An accurate number of electrons emitted over many periods reduces the noise.\cite{Kaestner2015,Albert2010,Maire2008} However, the low-frequency noise does not necessarily contain information about the regularity of the emitter. To characterize the regularity, it has been suggested to measure the distribution of electron waiting times between subsequent emission events.\cite{Brandes2008,Albert2011} For a highly regular emitter, the waiting time distribution (WTD) should be peaked around the period of the drive, corresponding to the emissions being separated in time exactly by the period.

\begin{figure}
\centering
\includegraphics[width=0.45\textwidth]{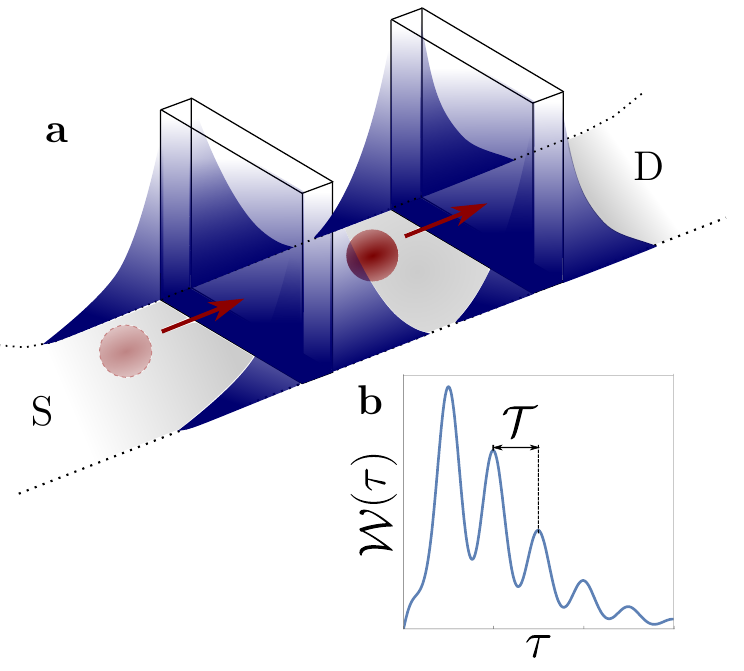}
\caption{(Color online) Dynamic single-electron turnstile and distribution of electron waiting times. \textbf{a,} The single-electron turnstile consists of a metallic island coupled to source
and  drain  electrodes  via  two  tunnel  junctions. Due to strong Coulomb interactions, the island can only be occupied by zero or one excess electrons. Using time-dependent gate voltages, the tunneling rates of the junctions are modulated periodically in time in order to regulate the single-electron transport. \textbf{b,} The waiting time between emitted electrons is denoted as~$\tau$. The distribution of waiting times $\mathcal{W}(\tau)$ is expected to be enhanced at multiplies of the period $\mathcal{T}$ of the drive.
}
\label{fig:Turnstile}
\end{figure}

In recent years, electron waiting times have been investigated theoretically for a variety of quantum transport setups. For Coulomb blockade structures such as quantum dots or metallic islands coupled to  normal-state or superconducting leads, methods based on Markovian\cite{Brandes2008,Albert2011} (and non-Markovian\cite{Thomas2013}) master equations have been developed. For coherent conductors, the distribution of electron waiting times can be obtained from a compact determinant formula containing the scattering matrix of the system \cite{Albert2012,Dasenbrook2014}. Moreover, transient behaviors have been described using non-equilibrium Green's functions.\cite{Tang2014,Tang2014Full,SeoaneSouto2015}

Based on these methods, electronic WTDs have been evaluated for a wide range of physical situations. A series of works have focused on WTDs of electron transport through single or double quantum dots\cite{Brandes2008,Welack2008,Welack2009,Welack2009Non,Thomas2013,Sothmann2014,Talbo2015,Rudge2016,Rudge20162,Ptaszynski2017}. Another line of research has been devoted to WTDs of mesoscopic conductors\cite{Albert2012,Haack2014}, including the influence of time-dependent perturbations \cite{Thomas2014,Albert2014,Dasenbrook2014,Dasenbrook2015,Hofer2016}. Distributions of waiting times have also been investigated for superconducting systems,\cite{Rajabi2013,Albert2016} for instance in relation to Josephson junctions\cite{Dambach2015,Dambach2016} and the detection of Majorana fermions.\cite{Chevallier2016} A theory of an electron waiting time clock has been developed\cite{Dasenbrook2016}, feedback control of electron waiting times has been proposed,\cite{Brandes2016} and connections between WTDs and quantum tomography have been identified.\cite{Haack2015}

The purpose of this paper is to develop a scheme for analytic calculations of the WTDs for periodically driven single-electron turnstiles, Fig.~\ref{fig:Turnstile}. Specifically, we use WTDs to understand the basic working principles of turnstiles and to characterize the regularity of the emission processes. In an earlier work, the WTD was evaluated for the special case of a single-electron emitter with a square-wave driving protocol at zero temperature.\cite{Albert2011} Here we present a method for calculating the WTD for arbitrary periodic driving protocols including finite temperature effects. Our method will be important for describing future experiments with arbitrary drivings and finite-temperature effects. We evaluate the distribution of electron waiting times for square-wave and harmonic driving protocols and we discuss in detail the crossover from adiabatic to non-adiabatic driving. Our predictions can readily be tested in future experiments on dynamic single-electron turnstiles using a capacitively coupled charge detector to measure the waiting times.\cite{Gustavsson2009}

The paper is organised as follows. In Sec.~\ref{sec:level3} we introduce the basic concepts of WTDs and the related idle-time probability with a specific focus on periodically driven emitters. In Sec.~\ref{sec:tunnel_junction} we illustrate these concepts by evaluating the distribution of electron waiting times for sequential tunneling through a driven tunnel junction.
In Sec.~\ref{sec:turnstile} we then go on to develop the theory of WTDs of periodically driven single-electron turnstiles. We introduce the rate equation description of the turnstile and show how to obtain the WTD for arbitrary driving protocols. We evaluate the periodic state of the emitter, the idle-time probability, and finally the WTD, going from the fully adiabatic to the strongly non-adiabatic regime. Finally, we discuss the influence of finite electronic temperatures. Our conclusions are presented in Sec.~\ref{sec:level5}.

\section{\label{sec:level3} Electron waiting times}

\begin{figure}
\centering
\includegraphics[width=0.45\textwidth]{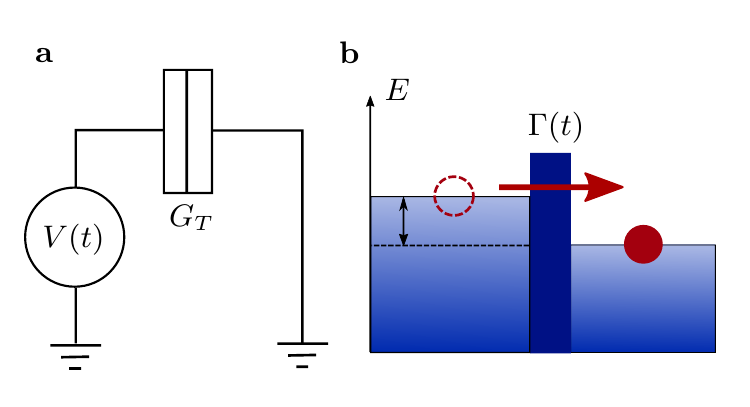}
\caption{(Color online) Dynamic tunnel junction. \textbf{a,} Circuit representation of the tunnel junction with tunnel conductance $G_T$ and a time-dependent voltage $V(t)$. \textbf{b,} Tunneling through the  junction occurs with the time-dependent rate $\Gamma(t)$ controlled by the applied voltage $V(t)$. The temperature is zero.}
\label{fig:single_barrier}
\end{figure}

The electron waiting time $\tau$ is the time that passes between two subsequent single-electron transfers through a nano-scale conductor.\cite{Brandes2008,Albert2011,Albert2012} Due to the stochastic nature of the charge transfer process, the electron waiting time is not a fixed quantity. Instead, it must be described by a distribution function $\mathcal{W}(\tau)$ which we refer to as the waiting time distribution (WTD). For stationary problems with no explicit time dependence, the WTD can be related to the idle-time probability $\Pi(\tau)$ as\cite{Albert2012,Haack2014}
\begin{equation}
    \mathcal{W} (\tau) = \langle \tau \rangle \partial_{\tau}^2 \Pi(\tau).
\label{eq:WTD}
\end{equation}
The idle-time probability $\Pi(\tau)$ is the probability that no electrons are transferred through the conductor during a time span of duration $\tau$. The mean waiting time can be expressed in terms of the idle-time probability as\cite{Albert2012,Haack2014}
\begin{equation}
    \langle \tau \rangle = \int\limits_0^{\infty} d \tau  \mathcal{W}(\tau)\tau=-\frac{1}{\dot{\Pi}(\tau=0)}.
\label{eq:aveWT}
\end{equation}
These relations are important, since it is often easier to calculate the idle-time probability and then obtain the WTD by differentiation.

\begin{figure*}
\centering
\includegraphics[width=0.98\textwidth]{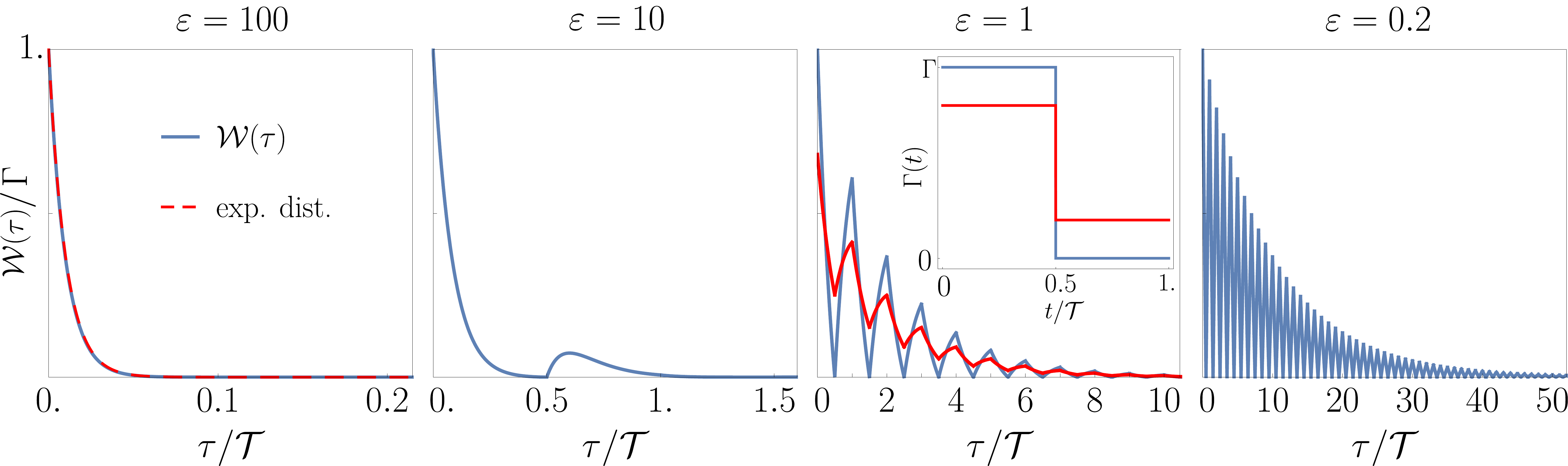}
\caption{(Color online) WTDs for a tunnel junction with square-wave driving. We show results for the adiabatic regime, where $\varepsilon=\Gamma\mathcal{T}\gg1$, to the non-adiabatic regime, $\varepsilon<1$. For $\varepsilon=1$, we consider WTDs for the two different  protocols in the inset [$\alpha=0$ (blue) and $\alpha=0.2$ (red)]. In the adiabatic regime, the WTD is well-approximated by an exponential distribution.}
\label{fig:WTDPoissStep}
\end{figure*}

In the following, we consider periodically driven single-electron emitters. In this case, the calculation of the WTD is complicated by the fact that the idle-time probability does not only depend on the length $\tau$ of the time interval $[t_0,t_0+\tau]$, but also on the initial time $t_0$.\cite{Dasenbrook2014,Dasenbrook2015,Hofer2016} The idle-time probability is then a two-time quantity that we denote as $\Pi(\tau,t_0)$. However, the relations above still hold, provided that we average the idle-time probability over a period of the drive $\mathcal{T}$ and define\cite{Dasenbrook2014,Dasenbrook2015,Hofer2016}
\begin{equation}
    \Pi(\tau) =  \frac{1}{\mathcal{T}}\int\limits_0^{\mathcal{T}} d t_0 \Pi(\tau, t_0).
    \label{eq:aveITP}
\end{equation}
In combination, Eqs.~(\ref{eq:WTD},\ref{eq:aveWT},\ref{eq:aveITP}) allow us to calculate the WTD for dynamically driven single-electron emitters. We now illustrate these ideas by evaluating the WTD for a dynamic tunnel junction before moving on to the more involved single-electron turnstile.

\section{Dynamic tunnel junction}
\label{sec:tunnel_junction}

We start by considering sequential tunneling through a single tunnel junction as illustrated in Fig.~\ref{fig:single_barrier}. To lowest order in the tunnel coupling, the rate for tunneling through the junction can be expressed as\cite{Averin1986,Pekola2013}
\begin{equation}
    \Gamma(t) = \frac{G_T}{e^2} \frac{\Delta E(t)}{e^{\beta \Delta E(t)}-1},
\label{eq:rate_finiteT}
\end{equation}
where $G_T$ is the tunneling conductance of the junction, $\beta=1/k_BT$ is the inverse temperature of the electronic leads, and $\Delta E(t)$ is the increase in energy due to a tunneling event. The tunneling rate takes into account the filled Fermi seas on both sides of the junction. For the tunnel junction, we have $\Delta E(t)=-eV(t)$, where $V(t)$ is the applied voltage. We focus here on voltage biases that are periodic in time such that $V(t+\mathcal{T})=V(t)$, where $\mathcal{T}$ is the period of the drive.  Higher-order tunneling processes are negligible, and we consider for now the zero temperature limit, where tunneling against the voltage does not occur. (Of course, in an experiment, the temperature will always be non-zero, but the zero temperature limit can still be a good approximation.) The tunneling rate is in this case proportional to the bias voltage
\begin{equation}
\begin{split}
    \Gamma(t) &= -\frac{G_T}{e^2} \Delta E(t)\Theta[-\Delta E(t)]\\
    &=\frac{G_T}{e}  V(t)\Theta[V(t)],\,\, T\rightarrow 0.
\end{split}
    \label{eq:rate_zeroT}
\end{equation}
Thus, by varying the voltage bias $V(t)$, we can control the time-dependence of the tunneling rate $\Gamma(t)$.

To evaluate the distribution of electron waiting times, it is useful to introduce the counting statistics of tunneling events described by the probability $P(n,t)$ of $n$ electrons having tunneled through the junction during the time span $[t_0,t_0+t]$.\cite{Bagrets2003,Flindt2008,Flindt2010} This probability evolves according to the rate equation
\begin{equation}
      \frac{d}{dt}P(n, t) = \Gamma(t) P(n-1, t) - \Gamma(t) P(n, t).
\label{eq:Pn_evolve}
\end{equation}
We moreover define the moment generating function
\begin{equation}
    \mathcal{M}(\chi,t) = \sum\limits_{n=0}^{\infty} P(n,t) e^{in\chi},
\label{eq:MGF}
\end{equation}
where $\chi$ is the counting field. The evolution of the moment generating function follows from Eq.~(\ref{eq:Pn_evolve}) and reads
\begin{equation}
      \frac{d}{dt}\mathcal{M}(\chi,t) = \Gamma(t) \left( e^{i \chi} -1 \right) \mathcal{M}(\chi,t).
\end{equation}
We then easily find
\begin{equation}
   \mathcal{M}(\chi,t) = e^{( e^{i \chi} -1)\int_{t_0}^{t_0+t} dt' \Gamma (t')} \mathcal{M}(\chi,t_0),
   \label{eq:MGF_sol}
\end{equation}
From the moment generating function we have access to all moments of $n$. However, we can also find the idle time probability. To this end, we note that\cite{Dasenbrook2016}
\begin{equation}
\mathcal{M}(i\infty,t)=P(n=0,t)
\end{equation}
which is exactly the idle time probability. From Eq.~(\ref{eq:MGF_sol}) we then find
\begin{equation}
    \Pi(\tau, t_0) = e^{-\int_{t_0}^{t_0+\tau} dt' \Gamma (t')} \Pi (0,t_0)
    \label{eq:ITP_single_junction}
\end{equation}
with the initial condition $\Pi (0,t_0)=1$. We note that this result could also have been reached by solving Eq.~(\ref{eq:Pn_evolve}) for $P(n=0,t)$ using that $P(n<0,t)=0$. However, when we consider the more involved turnstile in the following section, we will see that it is generally convenient to introduce a counting field as above.

\begin{figure*}
\centering
\includegraphics[width=0.98\textwidth]{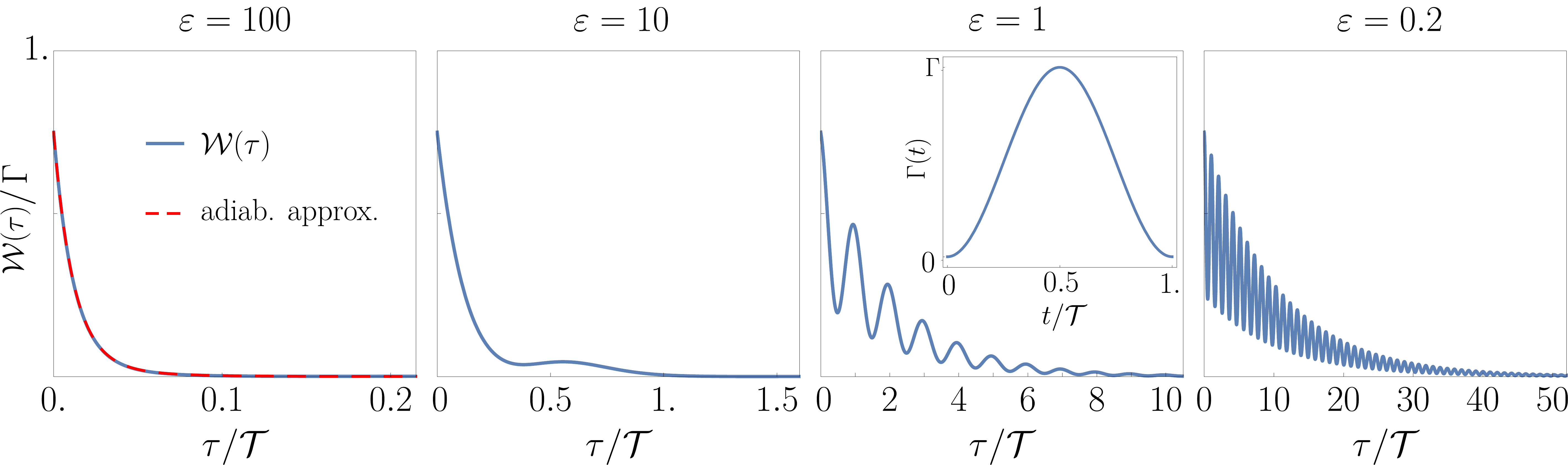}
\caption{(Color online) WTDs for a tunnel junction with harmonic driving.  We show results for the adiabatic regime, where $\varepsilon=\Gamma\mathcal{T}\gg1$, to the non-adiabatic regime, $\varepsilon<1$. In the adiabatic regime, the WTD is well-approximated by an average over Poisson processes with the instantaneous tunneling rate $\Gamma(t)$, see Eq.~(\ref{eq:poisapprox}) for the adiabatic approximation.}
\label{WTDPoissSine}
\end{figure*}

\subsection{Square-wave driving}

By combining Eq.~(\ref{eq:ITP_single_junction}) with Eqs.~(\ref{eq:WTD},\ref{eq:aveWT},\ref{eq:aveITP}) we can evaluate the distribution of electron waiting times for the tunnel junction. We start by considering the square-wave driving protocol
\begin{equation}\label{step}
    \Gamma(t) = \Gamma\left(1 - 2\alpha \right) \Theta \left(t-\floor*{t +\mathcal{T}/2}\right) + \alpha \Gamma,
\end{equation}
where $\floor{\cdot}$ denotes flooring, and the parameter $ \alpha\in[0,1/2]$ controls the amplitude of the drive. For $\alpha = 0$, the protocol is a periodic step-function with the rate $\Gamma$ in the on-state and the rate~0 in the off-state. For non-zero values of $\alpha$, the rate switches between $\Gamma(1 - \alpha)$ and  $\Gamma \alpha$. This may describe a leakage current in the off-state.

Carrying out the calculation of the WTD, we find for $\alpha = 0$ the compact result
\begin{equation}
    \mathcal{W} (\tau) = \Gamma e^{-\Gamma \tau} e^{\frac{\Gamma \mathcal{T}}{2} \floor*{ \frac{\tau + \mathcal{T}/2}{\mathcal{T}}}} \left|2 \left(  \floor*{ \frac{\tau}{\mathcal{T}}} -\frac{\tau}{\mathcal{T}}\right)+1  \right|.
\end{equation}
This result can be further simplified by introducing the dimensionless quantities
\begin{equation}
\begin{split}
s&=\tau/\mathcal{T},\\
\varepsilon &= \Gamma\mathcal{T},\\
\mathbb{W}(s)&=\mathcal{W}(\tau=s\mathcal{T})\mathcal{T},
\end{split}
\label{eq:dlessquant}
\end{equation}
leading to an appealing expression reading
\begin{equation}
    \mathbb{W}(s) = \varepsilon e^{-\varepsilon s } e^{\frac{\varepsilon}{2} \floor*{ s + \frac{1}{2}}} \left| 2\left( \floor*{s}-s \right)+1 \right|.
\end{equation}
Here we clearly see that the shape of the WTD is fully controlled by the dimensionless parameter $\varepsilon$ given by the tunneling rate $\Gamma$ times the period $\mathcal{T}$. Large values of  $\varepsilon$ correspond to the limit of slow (or adiabatic) driving, while small values of $\varepsilon$ describe non-adiabatic driving. In the adiabatic limit, most waiting times are short such that we can take $s\ll1 $ and approximate  $\mathbb{W}(s)\simeq \varepsilon e^{-\varepsilon s}$ corresponding to a Poisson process.

In Fig.~\ref{fig:WTDPoissStep} we show WTDs for the square-wave protocol. In the adiabatic limit, $\varepsilon\gg 1$, the WTD is essentially exponential with a mean waiting time which is much shorter than the period. A large number of electrons tunnel through the junction in the on-state, interrupted by quiet periods in the off-state. As the tunneling rate is decreased, a pronounced suppression of the WTD is found at $\tau=\mathcal{T}/2$, since electrons cannot tunnel with a separation in time of exactly half the period. As the inverse tunneling rate becomes comparable with the period, the WTD is suppressed to zero at times that are separated by the period of the drive. Additionally, the WTD is enhanced at multiples of the period. The suppression is partially lifted if the tunneling rate does not reach zero in the off-state. (The WTD for $\alpha\neq 0$ can be found in App.~\ref{AppA}). Finally, in the non-adiabatic regime, $\varepsilon< 1$, the synchronization between the drive and the tunneling is gradually lost.

\subsection{Harmonic driving}

Next, we consider the harmonic protocol
\begin{equation}\label{sin}
    \Gamma(t) = \Gamma \sin^2 \left(\pi t /\mathcal{T} \right).
\end{equation}
with period $\mathcal{T}$. In this case, we find for the WTD
\begin{widetext}
\begin{equation}\label{PoissSine}
\mathbb{W}(s) = e^{-\frac{\varepsilon s }{2}} \left( \frac{\varepsilon}{4} \mathcal{I}_0\left[ \frac{\varepsilon}{2 \pi} \sin \left( \pi s \right)\right] \left\{3 + \cos \left( 2 \pi s \right) \right\} - \mathcal{I}_1\left[ \frac{\varepsilon}{2 \pi} \sin \left( \pi s \right)\right] \left\{ \varepsilon \cos \left( \pi s \right) + \pi /\sin \left( \pi s \right) \right\} \right)
\end{equation}
\end{widetext}
in terms of the dimensionless quantities defined in Eq.~(\ref{eq:dlessquant}), and where $\mathcal{I}_0$ and $\mathcal{I}_1$ are zeroth and first order modified Bessel functions of the first kind.

The WTDs for the harmonic drive are shown in Fig.~\ref{WTDPoissSine}. In the adiabatic regime, $\varepsilon\gg 1$, the WTD is well-approximated by an average over Poisson processes with the instantaneous tunneling rate $\Gamma(t)$, such that
\begin{equation}
    \mathcal{W}(\tau) \simeq \frac{1}{\mathcal{T}}\int_0^{\mathcal{T}} d t \Gamma (t) e^{-\Gamma(t) \tau},\,\, \varepsilon\gg 1.
    \label{eq:poisapprox}
\end{equation}
As the tunneling rate is decreased, the WTDs start to develop oscillations due to the periodic drive, similar to the results in Fig.~\ref{fig:WTDPoissStep} for the square-wave driving. Again, in the non-adiabatic regime, the synchronization between the drive and the tunneling events is gradually lost.

\section{Single-electron turnstile}
\label{sec:turnstile}

\begin{figure}
\centering
\includegraphics[width=0.5\textwidth]{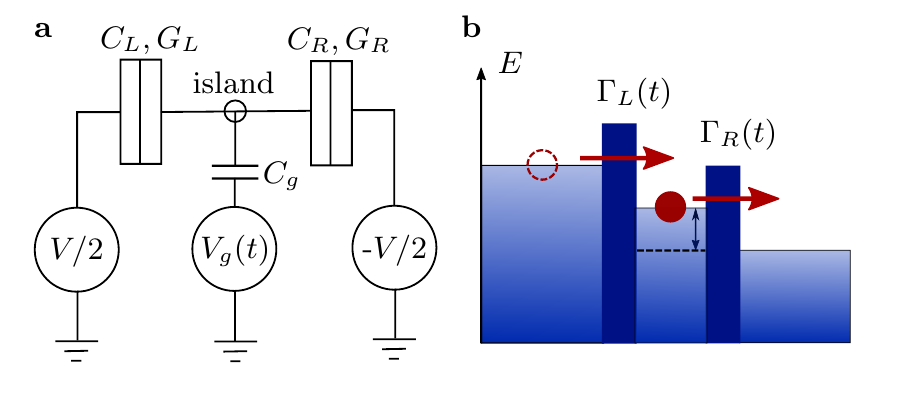}
\caption{(Color online) Dynamic single-electron turnstile. \textbf{a,} The turnstile consists of a metallic island coupled to source and drain electrodes via two tunnel junctions with capacitances $C_{L/R}$ and tunnel conductances $G_{L/R}$. A constant voltage $V$ ensures that the  transport is uni-directional at zero temperature. A time-dependent gate voltage $V_g(t)$ is used to modify the transport through the island. \textbf{b,} Tunneling through the tunnel junctions occurs with the time-dependent rates $\Gamma_L(t)$ and $\Gamma_R(t)$, controlled by the gate voltage $V_g(t)$.}
\label{turnstile}
\end{figure}

We are now ready to consider the single-electron turnstile depicted in Fig.~\ref{turnstile}. Unlike the tunnel junction from the previous section, we here need to keep track of the charge state of the turnstile. To this end, we consider a Markovian master equation of the form
\begin{equation}
    \frac{d}{dt} \vert p(t)\rangle = \mathbf{L}(t) \vert p(t)\rangle.
    \label{eq:master_eq}
\end{equation}
where $\vert p (t)\rangle$ is a column vector with the occupation probabilities for the different charge states of the turnstile and the matrix $\mathbf{L}(t)$ contains the time-dependent rates for making transitions between them. We use the compact bracket notation known from quantum mechanics, keeping in mind that we are considering an essentially classical transport process. As in the previous section, we introduce a counting field $\chi$ that couples to the number of electrons that have tunneled through the right junction. This is a standard procedure in full counting statistics,\cite{Bagrets2003,Flindt2008,Flindt2010} leading us to a modified master equation of the form
\begin{equation}
    \frac{d}{dt} \vert p_{\chi} (t)\rangle = \mathbf{L}_{\chi}(t) \vert p_{\chi} (t)\rangle.
    \label{eq:mod_master}
\end{equation}
For $\chi=0$, we recover the original master equation without the counting field. Below, we specify the rate matrix $\mathbf{L}_{\chi}(t)$ for the single-electron turnstile.  The modified master equation can formally be solved as
\begin{equation}
    \vert p_{\chi} (t)\rangle = \mathbf{U}_{\chi}(t,t_0) \vert p_{\chi} (t_0)\rangle,
\end{equation}
where the evolution operator is given by a time-ordered exponential as\cite{Pistolesi2004,Croy2016}
\begin{equation}
   \mathbf{U}_{\chi}(t,t_0) =  \widehat{T}\left\{ e^{\int_{t_0}^{t} dt' \mathbf{L}_{\chi}(t')}\right\}.
\end{equation}
In general, it is hard to evaluate the time-ordered exponential. However, for the single-electron turnstile, we can evaluate it for the particular values of the counting field that we need, specifically for $\chi=0$ and $\chi=i\infty$.

The moment generating function now reads
\begin{equation}
    \mathcal{M}(\chi,t) = \langle 1 \vert  \mathbf{U}_{\chi}(t,t_0) \vert p_{\chi}(t_0)\rangle,
\label{eq:MGF_turnstile}
\end{equation}
where $ \langle 1 \vert$ is a row vector of ones. Taking the limit $\chi\rightarrow i\infty$, we find the idle time probability as
\begin{equation}
    \Pi(\tau,t_0) = \langle 1 \vert  \mathbf{U}_{i\infty}(t_0+\tau,t_0) \vert p(t_0)\rangle,
\label{eq:ITP_turnstile}
\end{equation}
having used that $\vert p_{i\infty}(t_0)\rangle=\vert p(t_0)\rangle$ at the time $t_0$, when we start counting. For the initial state $\vert p(t_0)\rangle$, we assume that the turnstile has relaxed to its periodic state given by the normalized solution to the equation
\begin{equation}
    \vert p(t_0)\rangle = \mathbf{U}_{0}(t_0+\mathcal{T},t_0) \vert p(t_0)\rangle
\label{eq:periodic_state}
\end{equation}
with $\vert p(t_0+\mathcal{T})\rangle=\vert p(t_0)\rangle$ by definition. Combined with Eqs.~(\ref{eq:WTD},\ref{eq:aveWT},\ref{eq:aveITP}) we can then evaluate the distribution of electron waiting times for the single-electron turnstile.

\begin{figure*}
\centering
\includegraphics[width=0.8\textwidth]{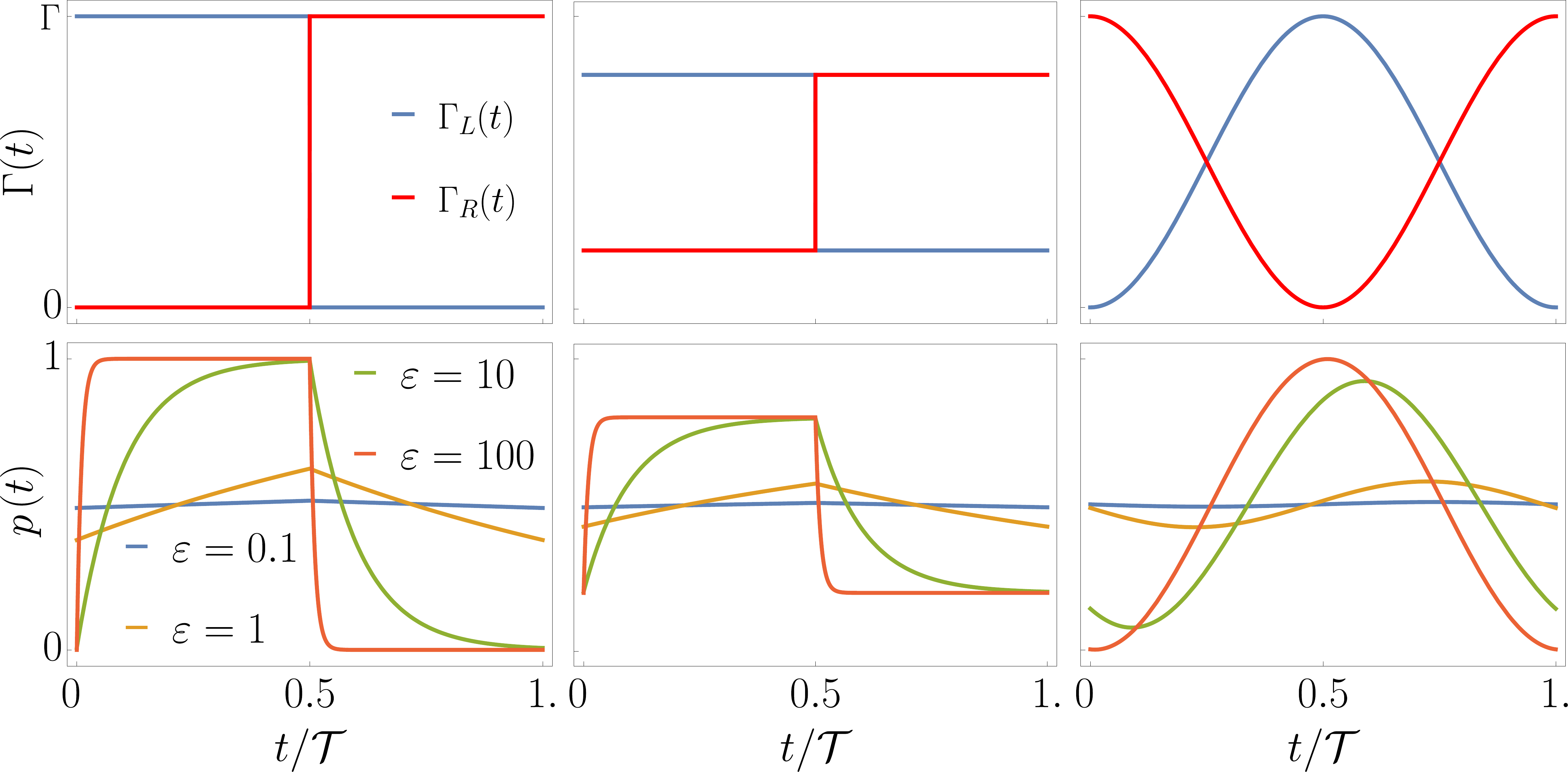}
\caption{(Color online) Driving protocols and occupation probabilities. In the upper row, we show three different driving protocols. The lower row shows the corresponding occupation probabilities for different values of $\varepsilon=\Gamma\mathcal{T}$.}
\label{p1}
\end{figure*}

\subsection{Master equation}

Next, we specify the rate matrix for the turnstile. The turnstile consists of a metallic island coupled via tunnel junctions to a source and a drain electrode as illustrated in Fig.~\ref{turnstile}. The island is operated close to a charge degeneracy point, where strong Coulomb interactions restrict the number of excess electrons on the island to zero or one. An applied voltage bias $V$ ensures that the electron transport is unidirectional from the source to the drain via the island. A time-dependent gate voltage $V_g(t)=V_g(t+\mathcal{T})$ is used to modulate the transport through the turnstile periodically in time. Again, we first consider the system at zero temperature for which the tunneling rates through the tunnel junctions read\cite{Averin1986,Pekola2013}
\begin{equation}
    \Gamma_{L}(t) =
         -\frac{G_L}{e^2} \Delta E_{L} (t)\Theta[-\Delta E_{L} (t)]
\end{equation}
with
\begin{equation}
    \Delta E_{L} (t) = E_{c} \left[1 - 2\left\{ C_gV_g(t) +  (C_g/2+C_L) V\right\}/e \right],
    \nonumber
\end{equation}
and
\begin{equation}
    \Gamma_{R}(t) =
         -\frac{G_R}{e^2} \Delta E_{R} (t)\Theta[-\Delta E_{R} (t)],
\end{equation}
with
\begin{equation}
    \Delta E_{R} (t) = -E_{c} \left[1 -2\{C_gV_g(t) -  (C_g/2+C_R)V\}/e \right].
    \nonumber
\end{equation}
Here, $G_L$ and $G_R$ are the tunnel conductances of each junction, and the charging energy $E_c=e^2/[2(C_g+C_L+C_R)]$ of the island is expressed in terms of the junction and gate capacitances. It is convenient to consider identical tunnel junctions, $G_L=G_R=G_T$, so that
\begin{equation}
     \Gamma_L(t) + \Gamma_R(t)= G_TV/e=\Gamma
\end{equation}
is constant. For the single-electron turnstile, we see that we can modulate the individual tunneling rates in time using the gate voltage $V_g(t)$, while the overall amplitude $\Gamma$ can be controlled by the applied voltage bias $V$.

Since the island only has two charge states (empty or occupied), the rate matrix takes the simple form
\begin{equation} \label{rateeq}
\mathbf{L}_{\chi}(t) = \left( \begin{array}{cc}
     -\Gamma_L(t) & \Gamma_R(t) e^{i \chi}\\\\
     \Gamma_L(t) & -\Gamma_R(t)
\end{array}
\right),
\end{equation}
where we have included the counting factor $e^{i \chi}$ in the upper off-diagonal element together with $\Gamma_R(t)$, corresponding to counting the number of electrons that have tunneled through the right junction.\cite{Bagrets2003,Flindt2008,Flindt2010} We note that this particular form of the rate matrix is not restricted to metallic islands only, but it can also be used to describe transport through single-level quantum dots for example.

\subsection{Periodic state}

To evaluate the WTD, we need the periodic state of the turnstile, defined by the requirement that  $\vert p(t+\mathcal{T})\rangle=\vert p(t)\rangle$. To this end, we note that the probabilities for the island to be empty or occupied by a single electron must sum to one, i.e.~$p_0(t) + p_1(t) = 1$. We can then work with just the probability of the island to be occupied and write $p_1(t)=p(t)$ and $p_0(t)=1-p(t)$. In this case, the master equation in Eq.~(\ref{eq:master_eq}) can be converted into an ordinary differential equation for $p(t)$ reading
\begin{equation}
\frac{d}{dt}p(t)= \Gamma_L(t)-[\Gamma_L(t)+\Gamma_R(t)]p(t).
\end{equation}
Imposing the condition $p(t+\mathcal{T})=p(t)$, we then find
\begin{equation}
p(t)= \frac{e^{ -\int_{t}^{t+\mathcal{T}} dt'\sum_{\gamma} \Gamma_{\gamma}(t')  } \int_{t}^{t+\mathcal{T}} dt' \Gamma_L(t') e^{ \int_{t}^{t'} dt''\sum\limits_{\gamma} \Gamma_{\gamma}(t'') }}{1 - e^{ -\int_{t}^{t + \mathcal{T}}dt' \sum_{\gamma} \Gamma_{\gamma}(t')  }},
\end{equation}
where the sums run over the two junctions, $\gamma=L,R$.

The periodic state can be found for arbitrary periodic driving protocols. As examples, we consider square-wave and harmonic protocols. For the square-wave driving, the tunneling rates read
\begin{equation}
\begin{split}
\Gamma_L(t) &= \Gamma\Theta \left(t-\floor*{t+\mathcal{T}/2}\right),\\
\Gamma_R(t) &= \Gamma\left[1-\Theta \left(t-\floor*{t+\mathcal{T}/2}\right)\right].
\end{split}
    \label{eq:turnstep}
\end{equation}
In this case, we find for the occupation probability
\begin{equation}
    p(t) = \left\{ \begin{array}{cl}
  1 + e^{-\Gamma t} \frac{\left( e^{-\Gamma\mathcal{T}/2} -1 \right)}{\left(1 - e^{-\Gamma\mathcal{T}} \right)},& 0 \leq t < \frac{\mathcal{T}}{2} \\\\
   e^{-\Gamma t} \frac{\left( e^{-\Gamma\mathcal{T}/2} -1 \right)}{\left(1 - e^{-\Gamma\mathcal{T}} \right)},&  \frac{\mathcal{T}}{2} \leq t \leq \mathcal{T} \\
\end{array} \right.
\end{equation}
which is repeated with the period $\mathcal{T}$. For the harmonic drive, we take
\begin{equation} \label{turnsin}
\begin{split}
    \Gamma_L(t) & = \Gamma \sin^2 \left(\pi t /\mathcal{T} \right),\\
   \Gamma_R(t) &= \Gamma \cos^2 \left(\pi t /\mathcal{T} \right),
    \end{split}
    \end{equation}
and find for the occupation probability
\begin{equation}
    p(t) = \frac{1}{2} \left(1 - \frac{\cos(2\pi t /\mathcal{T} ) + \frac{2 \pi}{\Gamma\mathcal{T}}  \sin(2\pi t /\mathcal{T} )}{1-\left( \frac{2 \pi}{\Gamma\mathcal{T}} \right)^2}    \right).
\end{equation}

\begin{figure*}
\centering
\includegraphics[width=0.98\textwidth]{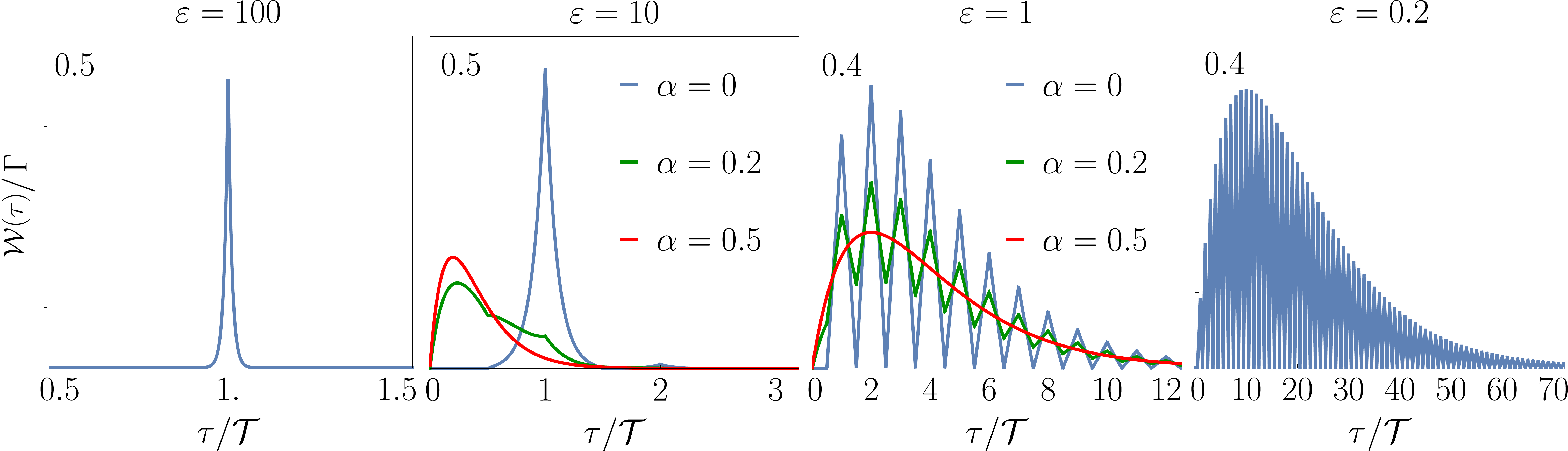}
\caption{(Color online) Distribution of electron waiting times for a single-electron turnstile driven by a square-wave gate voltage. The rates switch periodically between $\Gamma(1-\alpha)$ and $\Gamma \alpha$ with $\alpha=0$ (blue curves), $\alpha=0.2$ (green curves), and $\alpha=0.5$ (red curves).  The results cover the transition from the adiabatic regime, where $\varepsilon=\Gamma\mathcal{T}\gg1$, to the non-adiabatic regime, $\varepsilon<1$.}
\label{fig:WTDturn_step}
\end{figure*}

In Fig.~\ref{p1}, we show the driving protocols together with the occupation probabilities. In the adiabatic regime, $\varepsilon=\Gamma\mathcal{T}\gg1$, the island quickly responds to the change of the tunneling rates, and the occupation probability closely follows the rate for tunneling through the left junction. As the tunneling rates are decreased, the occupation probability starts to lack behind the drive and a clear retardation effect is observed. Finally, in the non-adiabatic regime, $\varepsilon<1$, the synchronization with the drive is gradually lost, and the occupation probability becomes nearly constant.

\subsection{Idle-time probability}

With the periodic state at hand, we can calculate the idle-time probability. Here, we need the time evolution operator evaluated in the limit $\chi\rightarrow i\infty$. In this limit, the upper off-diagonal element of the matrix in Eq.~(\ref{rateeq}) vanishes, and the time evolution operator takes the form
\begin{equation}
\mathbf{U}_{i\infty} (t, t_0) = \left( \begin{array}{cc}
    U_{i\infty}^{11}(t,t_0) & 0\\
    U_{i\infty}^{21} (t,t_0) & U_{i\infty}^{22} (t,t_0)
\end{array}
\right)
\end{equation}
with the non-zero elements reading
\begin{equation}
 \begin{split}
    U_{i\infty}^{11}(t,t_0)  &= e^{ -\int_{t_0}^{t} dt' \Gamma_L(t')  },\\
    U_{i\infty}^{22}(t,t_0)  &= e^{ -\int_{t_0}^{t} dt'\Gamma_R(t')  },
\end{split}
\end{equation}
and
\begin{equation}
 \begin{split}
        U_{i\infty}^{21}(t,t_0)&= e^{ -\int_{t_0}^{t}dt' \Gamma_R(t') } \\
        &\times\int_{t_0}^{t}dt' \Gamma_L(t') e^{ \int_{t_0}^{t'}dt'' \left[\Gamma_{R} (t'')-\Gamma_{L} (t'') \right] }.
        \end{split}
\end{equation}
The idle-time probability can then be written as
\begin{equation}
 \begin{split}
        \Pi(\tau,t_0) & =  U_{i\infty}^{11}(t_0+\tau,t_0)[1-p (t_0)]\\
        &+ U_{i\infty}^{21}(t_0+\tau,t_0)[1-p (t_0)]\\
        &+ U_{i\infty}^{22}(t_0+\tau,t_0) p (t_0)
        \end{split}
\end{equation}
allowing us to evaluate the distribution of waiting times.

\subsection{Square-wave driving}

By combining Eqs.~(\ref{eq:WTD},\ref{eq:aveWT},\ref{eq:aveITP}) with the idle-time probability above, we can find the WTD for the turnstile. For the square-wave driving we find the compact result
\begin{equation}
    \mathbb{W} (s) = \varepsilon\floor*{s+ 1/2}  e^{-\varepsilon \floor*{s + 1/2}/2} \sinh \left| \varepsilon \left( \floor*{s}-s+1/2\right) \right|,
    \nonumber
\end{equation}
which previously has been derived in Ref.~\onlinecite{Albert2011}, however, without using the general method developed here. In addition, we can evaluate the distribution of electron waiting times in the case, where the rates switch periodically between the values $\Gamma(1-\alpha)$ and $\Gamma \alpha$ for $\alpha\in[0,1/2]$.

In Fig.~\ref{fig:WTDturn_step}, we show WTDs for the square-wave driving protocol. In the adiabatic regime, $\varepsilon=\Gamma\mathcal{T}\gg1$, the WTD is strongly peaked around the period of the drive. In this case, the emission of electrons is highly regular with essentially one electron being emitter in each period. The width of the peak is due to the uncertainty in the exact emission time of each electron. As the tunneling rates are lowered, the width of the peak increases and additional peaks appear at multiplies of the period. These peaks are due to cycle-missing events in which the turnstile fails to emit an electron within a period. One may then have to wait several periods between emission events. Finally, in the non-adiabatic regime, $\varepsilon<1$, the synchronization with the drive is gradually lost. We note that two emission events can never be separated by less than half a period, implying that the WTD is suppressed to zero for $\tau\leq \mathcal{T}/2$ for all values of $\varepsilon$.

In Fig.~\ref{fig:WTDturn_step}, we also show results for the case where the tunneling rates do not reach zero in the off-state. The analytic expression for the WTD with $\alpha\neq0$ is lengthy and is not shown here. As the parameter $\alpha$ is tuned from 0 to 1/2, the WTD approaches the result for two static tunnel barriers in series\cite{Brandes2008}
\begin{equation}
\mathcal{W}(\tau)=
\frac{\Gamma_L\Gamma_R}{\Gamma_L-\Gamma_R} \left( e^{-\Gamma_R \tau} - e^{-\Gamma_L \tau} \right)
\label{eq:stat_WTD}
\end{equation}
with $\Gamma_L=\Gamma_R=\Gamma/2$, such that $\mathcal{W}(\tau)=(\Gamma/2)^2\tau e^{-\Gamma\tau/2}$.

\begin{figure*}
\centering
\includegraphics[width=0.98\textwidth]{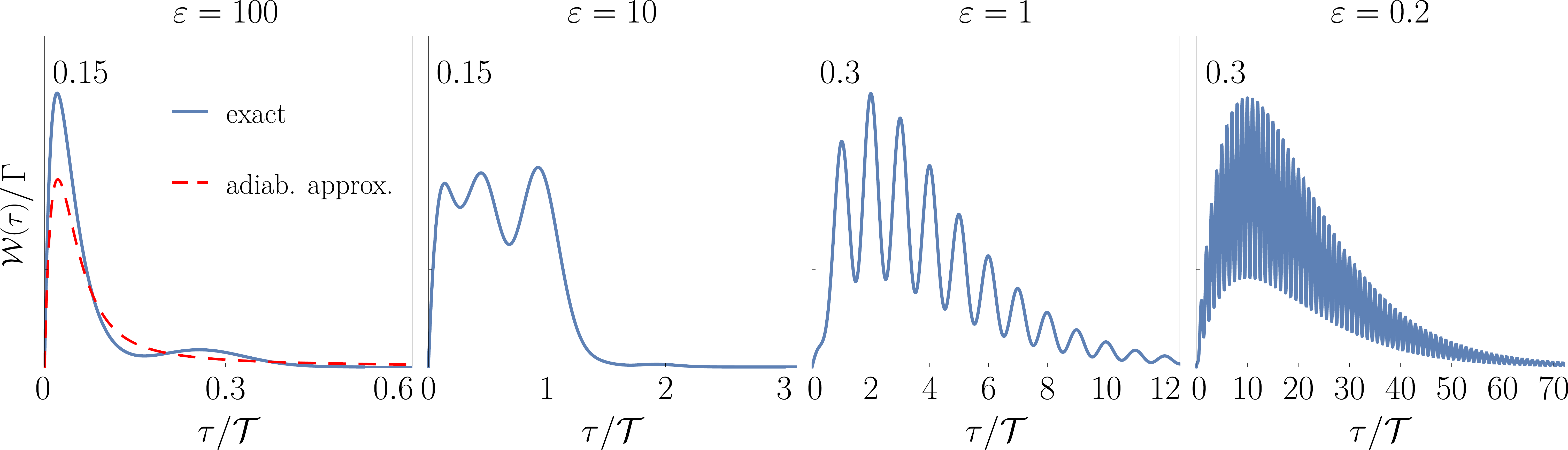}
\caption{(Color online) Distribution of electron waiting times for a single-electron turnstile driven by a harmonic gate voltage. We show results for the adiabatic regime, where $\varepsilon=\Gamma\mathcal{T}\gg1$, to the non-adiabatic regime, $\varepsilon<1$. In the adiabatic regime, the WTD can be approximated by an average over WTDs corresponding to stationary processes with fixed rates, see Eq.~(\ref{eq:averaged_WTD}).}
\label{fig:WTDturn_sin}
\end{figure*}

\subsection{Harmonic driving}

For the harmonic drive, the elements of the time evolution operator entering the idle-time probability read
\begin{equation}
\begin{split}
    U_{i\infty}^{11}(t_0+\tau,t_0)  &= e^{-\frac{\Gamma \mathcal{T} }{2 \pi} \left(  \frac{ \pi}{\mathcal{T}}\tau  -  \cos\left[ \frac{\pi}{\mathcal{T}} \tau + \frac{2 \pi}{\mathcal{T}} t_0 \right] \sin\left[ \frac{\pi}{\mathcal{T}} \tau \right] \right)},\\
    U_{i\infty}^{22}(t_0+\tau,t_0)  &= e^{-\frac{\Gamma \mathcal{T} }{2 \pi} \left( \frac{ \pi}{\mathcal{T}}\tau  + \cos\left[ \frac{\pi}{\mathcal{T}} \tau + \frac{2 \pi}{\mathcal{T}} t_0 \right] \sin\left[ \frac{\pi}{\mathcal{T}} \tau \right] \right)},
\end{split}
\end{equation}
and
\begin{equation}
\begin{split}
    U_{i\infty}^{21}(t_0+\tau,t_0)  = \frac{1}{2} e^{-\frac{\Gamma \mathcal{T} }{2 \pi} \left( \frac{ \pi}{\mathcal{T}}\tau  + \cos\left[ \frac{\pi}{\mathcal{T}} \tau + \frac{2 \pi}{\mathcal{T}} t_0 \right] \sin\left[ \frac{\pi}{\mathcal{T}} \tau \right]\right)}\\
    \times \Big[ \left(1-e^{\frac{\Gamma \mathcal{T} }{2 \pi} \left( \sin\left[ \frac{2 \pi}{\mathcal{T}} (t_0 + \tau) \right]-\sin\left[ \frac{2 \pi}{\mathcal{T}} t_0  \right]\right)} \right) \\
    + \Gamma e^{-\frac{\Gamma \mathcal{T} }{2 \pi} \sin\left[ \frac{2 \pi}{\mathcal{T}} t_0 \right]} \int\limits_{t_0}^{t_0+\tau} dt  e^{\frac{\Gamma \mathcal{T} }{2 \pi} \sin\left[ \frac{2 \pi}{\mathcal{T}} t \right]} \Big].
    \end{split}
    \label{eq:U21}
\end{equation}
To proceed, we expand the integrand above as
\begin{equation}
e^{\frac{\Gamma \mathcal{T} }{2 \pi} \sin\left[ \frac{2 \pi}{\mathcal{T}} t \right]} \simeq 1+\frac{\Gamma \mathcal{T} }{2 \pi} \sin\left[ \frac{2 \pi}{\mathcal{T}} t \right]+\ldots,
\end{equation}
allowing us to evaluate the integral in Eq.~(\ref{eq:U21}) order by order in $\varepsilon=\Gamma \mathcal{T}$ for $\varepsilon\lesssim 2\pi$. The resulting expression for the WTD to second order in $\varepsilon$ agrees well with numerical results in the appropriate parameter range. Again, the analytic expression is lengthy and not shown here. For  $\varepsilon> 2\pi$, we evaluate the WTD numerically.

In Fig.~\ref{fig:WTDturn_sin}, we show WTDs for the harmonic driving protocol. In the adiabatic regime, $\varepsilon\gg1$, the WTD can be approximated by a time-average over WTDs for two static tunnel barriers in series as
\begin{equation}
    \mathcal{W}(\tau) = \int\limits_{0}^{\mathcal{T}} \frac{dt_0}{\mathcal{T}}  \mathcal{W}_{t_0} (\tau),
    \label{eq:averaged_WTD}
\end{equation}
where $\mathcal{W}_{t_0} (\tau)$ is given by Eq.~(\ref{eq:stat_WTD}) and the subscript indicates that we should use the tunneling rates $\Gamma_L(t_0)$ and $\Gamma_R(t_0)$ at the time $t_0$. Unlike the square-wave drive, the harmonic protocol does not lead to regular emission of single electrons separated by the period of the drive. At each instant of time, electrons can both enter and leave the island. For this reason, the harmonic driving is less efficient in regulating the electron transport. As the tunneling rate is lowered, the WTD starts to develop a peak at the period of the drive. However, cycle-missing events quickly become dominating, and peaks appear at multiplies of the period. Finally, in the non-adiabatic regime, the synchronization with the drive is gradually lost.

\subsection{Finite electronic temperatures}

So far, we have analyzed the zero-temperature limit. We now consider finite electronic temperatures. In this case, the tunneling rates read
\begin{equation}
    \Gamma_{\alpha}^{(\pm)} (t) = \frac{G_\alpha}{e^2} \frac{\Delta E_\alpha^{(\pm)} (t)}{e^{\beta \Delta E_\alpha^{(\pm)} (t)}-1},
\end{equation}
where $\Delta E_\alpha^{(\pm)} (t)$ is the increase in energy due to adding/removing ($\pm$) an electron to/from the island by tunneling through the left/right junction with tunnel conductance $G_\alpha$, $\alpha=L,R$. Due to the finite  temperature, electrons can now tunnel against the  bias. The modified rate matrix then takes the form
\begin{equation}
  \mathbf{L}_{\chi}(t)=  \left( \begin{array}{cc}
     -\Gamma_L^{(+)}(t) - \Gamma_R^{(+)}(t)  & \Gamma_L^{(-)}(t) + \Gamma_R^{(-)}(t) e^{i \chi}\\\\
     \Gamma_L^{(+)}(t) + \Gamma_R^{(+)}(t) & -\Gamma_L^{(-)}(t) - \Gamma_R^{(-)}(t)
\end{array}
\right),
\end{equation}
where we again have added a counting field that couples to the number of electrons that have tunneled from the island to the right lead. This choice of the counting field corresponds to measuring the waiting time between electrons emitted into the drain, while disregarding those that are absorbed. In this case, we are not able to calculate the idle-time probability analytically. Instead, we solve Eq.~(\ref{eq:mod_master}) numerically in the limit $\chi \rightarrow i \infty$ and then find the idle-time probability according to Eq.~(\ref{eq:ITP_turnstile}).

\begin{figure}
\centering
\includegraphics[width=0.95\columnwidth]{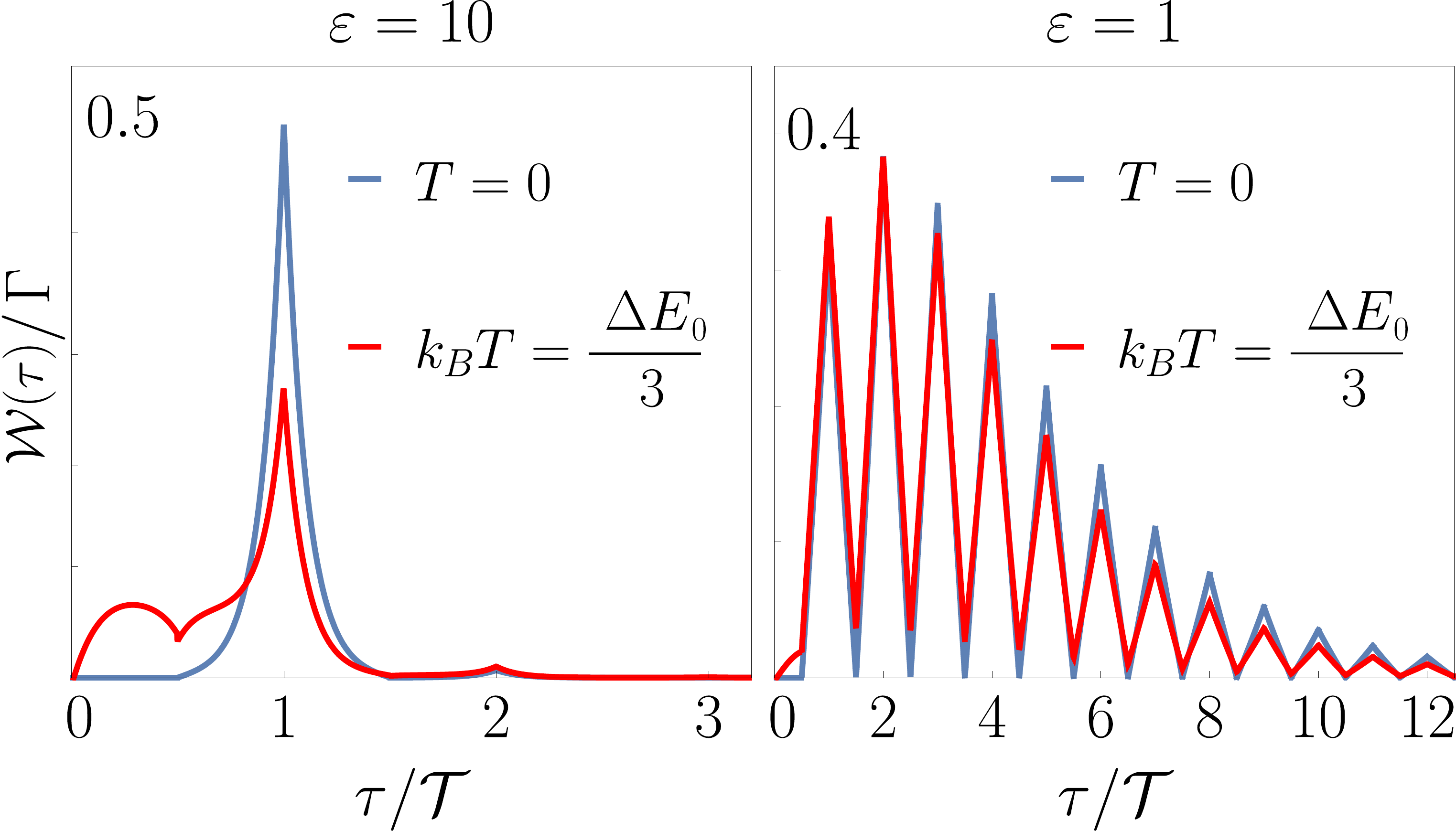}
\caption{(Color online) Distribution of electron waiting times for finite electronic temperatures. Results are shown for the square-wave protocol with $\varepsilon=\Gamma\mathcal{T}=1$ and $\varepsilon=10$. Temperatures are $T=0$ (blue) and $T=\Delta E_0/(3k_B)$ (red), where $\Delta E_0$ is the maximal value of $|\Delta E^{(\pm)}_\alpha(t)|$ during the protocol.}
\label{WTDturnT}
\end{figure}

The effect of finite electronic temperatures can be seen in Fig.~\ref{WTDturnT}. Here, we compare WTDs for the square-wave driving protocol at zero and at finite temperatures. In the adiabatic regime, the finite electronic temperature degrades the regularity of the single-electron emitter as it allows for the island to be refilled (emptied) during the unloading (loading) phase. This leads to less regular emissions of electrons which are not separated by the period of the drive. As we approach the non-adiabatic regime, the influence of a finite electronic temperature is less dramatic. Still, we see that the suppression of the WTD to zero is lifted by the finite electronic temperature.

\section{\label{sec:level5}Conclusions}

We have investigated the distribution of waiting times between electrons emitted from a periodically driven single-electron turnstile. To this end, we have a developed a general scheme for analytic calculations of the WTD for arbitrary periodic driving protocols. Our method will be important for describing future experiments with arbitrary drivings and finite-temperature effects. The WTDs provide us with clear insights into the single-electron emission processes from the driven turnstile and their regularity. This information is complementary
to what can be learned from conventional current measurements. Our predictions can be  tested in future  experiments on dynamic single-electron turnstiles using a capacitively coupled charge detector to measure the distribution of electron waiting times.

\section{\label{sec:ackn}Acknowledgements}
We dedicate this paper to the memory of Tobias Brandes. We thank Michael Moskalets for useful comments on the manuscript. Both authors are affiliated with Centre for Quantum Engineering at Aalto University.

\appendix

\section{\label{AppA}WTD for the tunnel junction}

For the single tunnel junction driven by square wave pulses, we find for $ \alpha\in[0,1/2]$ the general result
\begin{widetext}
\begin{equation}
\begin{split}
    \mathbb{W}(s) = \frac{1}{ \left( 1 - 2 \alpha \right)} &\Big| e^{ (\alpha-1) s \varepsilon} e^{\left( \frac{1}{2}-\alpha \right) \varepsilon \floor*{s+ \frac{1}{2}}} \left(1 - \alpha \right)
    \Big[2 s \varepsilon \left( \alpha -1 \right)\left( 2\alpha -1 \right) + 4 \alpha + \varepsilon \left(2  \floor*{s} +1  \right)
 \left( \alpha (3-2\alpha)-1 \right) \Big]\\ &
 -  e^{-\alpha s \varepsilon} e^{-\left( \frac{1}{2}-\alpha \right) \varepsilon \floor*{s+ \frac{ 1}{2}}}
    \times \Big[4\alpha +\alpha^2 \left( \varepsilon \left( 2 s -\left( 2  \floor*{s} +1 \right) \left( 2 \alpha-1 \right) \right) -4 \right)  \Big] \Big|.
    \end{split}
\end{equation}
\end{widetext}

\end{document}